\documentclass[preprint, aps,prd,amsmath,amssymb, superscriptaddress, twoside, nofootinbib, showkeys, floatfix, longbibliography]{revtex4-2}

\usepackage{orcidlink}
\usepackage{physics,relsize}
\usepackage[e]{esvect}
\usepackage{bbm,comment}
\usepackage{hyperref}
\hypersetup{
    colorlinks=true,
    linkcolor=blue,
    filecolor=magenta,      
    urlcolor=cyan,
    pdftitle={Overleaf Example},
    pdfpagemode=FullScreen,
    }
\DeclareMathOperator{\diag}{diag}
\newcommand{\overbar}[1]{\mkern 1.5mu\overline{\mkern-1.5mu#1\mkern-1.5mu}\mkern 1.5mu}
\begin{document}
\title{Tensorial charge assignments in unitary groups}

\author{E. Castillo-Ruiz \orcidlink{0000-0001-9713-6040}}
\email{ocastillo@uni.edu.pe}

\author{Henry Diaz \orcidlink{0000-0001-6110-6130}}
\email{hdiaz@uni.edu.pe}
\affiliation{Facultad de Ciencias, Universidad Nacional de Ingeniería (UNI)\\ Av. Túpac Amaru s/n, Lima - Rímac 15333, Perú.}

\author{V. Pleitez \orcidlink{0000-0001-5279-8438}}
\email{v.pleitez@unesp.br}
\affiliation{Instituto de Física Teórica, Universidade Estadual Paulista UNESP \\ 01140-070, Rua Dr. Bento Teobaldo Ferraz 271, Barra Funda, S\~ao Paulo-SP,  Brazil.}

\date{\today}

\begin{abstract}
We present an index-based tensorial formulation for computing eigenvalues of charge operators acting on arbitrary tensor representations of unitary gauge groups. The construction follows directly from the action of Cartan generators on tensor products and the additivity of weights, leading to a compact operator acting on general \((i_p,i_q)\) tensors. This framework provides a practical bookkeeping tool for assigning charges to arbitrary-dimensional multiplets appearing in model building. Explicit applications to \(SU(2)\), \(SU(3)\), and \(SU(5)\) representations are discussed.
\end{abstract}

\maketitle

\section{Introduction}
\label{sec:intro}

Assigning conserved charges to multiplets transforming under gauge symmetries is a basic aspect of quantum field theory and particle physics. In particular, conservation laws linked to $U(1)$ factors arise very often, since in gauge theories, every conserved charge corresponds to specific abelian (diagonal) generators, which we analyze in a Cartan subalgebra. In practice, charge eigenvalues are commonly obtained by decomposing representations into weight (eigenstates of Cartan generators) and/or by using Young-diagram and Young-tableaux techniques to organize tensor representations and their decompositions; see, e.g., \cite{Slansky1981} for pedagogical summaries and applications to particle-physics model building. These methods are well established in the representation theory of Lie algebras.

While these representation-theoretic tools are conceptually clear, their direct application to fields transforming as higher-rank tensors can become cumbersome in explicit model-building contexts. In particular, when dealing with exotic multiplets in tensor representations involving both fundamental and antifundamental indices, it is often desirable to have a compact and systematic prescription for determining charge assignments without performing an explicit decomposition into irreducible components, characterizing possible physical multiplets without taking into account which fields they are connected to via the Lagrangian.

The purpose of this work is to provide a prescription in the form of an index-based tensorial formulation of charge operators. The construction presented here does not introduce new charge quantization principles. Rather, it reformulates known representation-theoretic results—namely, the action of Cartan generators on tensor products and the additivity of weights—into a single operator acting directly on general tensors of type $(i_p,j_q)\subset V^{\otimes p}\otimes (V^*)^{\otimes q}$ with components $T^{i_1\cdots i_p}{}_{j_1\cdots j_q}$. This formulation allows charge eigenvalues to be derived by simple index bookkeeping rules.

For example, with just an upper index, the electric charge tensor $Q^i$ can be defined in matrix form by the Gell-Mann-Nishijima (GMN) formula int the case of $SU(2)$ \cite{Gell-Mann:1956iqa,10.1143/PTP.13.285}. 
\begin{equation}
Q=T_3+\frac{Y}{2}\mathbbm{1},
\label{gnf}
\end{equation}
which stands as a cornerstone in particle physics, providing a fundamental connection between a particle's internal quantum numbers and its electric charge. However, to explore scenarios beyond the Standard Model, such as exotic particles with unconventional charges or models with extended symmetries, a more general framework than Eq.~\eqref{gnf} is needed.

Once we have handled $n$-dimensional representations of $SU(n)$, the electric charge assignment follows directly, provided the diagonal generators are known explicitly. As an illustration, in $SU(2)\otimes U(1)_Y$ we employ the expression in Eq.~\eqref{gnf} for the fundamental representation; however, if we wish to build the adjoint representation, we must instead use the triplet generator $T_3 = \textrm{Diag}(+1,\,0,\,-1)$. This implies that, for example, when $Y=0$, we obtain the triplet $(\phi^+\,\phi^0\,\phi^-)$. The $\phi$ fields may correspond to scalars, fermions, or vectors. More generally, for $SU(n)$ it suffices to know its $T_3$ generator. Although Eq.~\eqref{gnf} has already appeared in the strong-isospin framework, it is equally applicable in the context of electroweak or grand unified theories (see early references in \cite{Rosner:2002xi}), considering the abelian generators of the given group, even within larger semi-simple groups. In such scenarios, the electric charges of exotic multiplets are sometimes determined by analyzing how new particles interact with known ones, but in every case, the charge values can be interpreted as weights of the Cartan generators.

In this work, we construct the GMN formula for representations of arbitrary dimension, with charge assignments determined directly by intrinsic group properties rather than specific particle interactions, enabling the description of dark-matter candidates and models with horizontal, color, or other symmetries. In this framework, Eq.~\eqref{gnf} emerges as the lowest nontrivial two-dimensional case, with $Q^i, i=1,2$ in $SU(2)\otimes U(1)_Y$, and more generally, we define a tensor $Q^{i_1 i_2 \ldots}_{j_1 j_2 \ldots}$ for any symmetry group, applicable to both local and global theories. Starting from a generic charge operator written as a linear combination of mutually commuting generators, we derive a closed-form expression for its action on arbitrary tensor representations, capturing in explicit index notation the additive action of generators on tensor products and the sign flip for dual representations, thereby linking the usual weight-space formalism to a concrete tensor-level realization suitable for specifying Lagrangian field content. We then apply this general formalism to standard unitary groups and representations, including $SU(2)$, $SU(3)$, and $SU(5)$, where the examples do not introduce new charge assignments but instead show how the tensor-index formulation systematically reproduces known results and can be efficiently extended to higher-rank or less familiar multiplets in phenomenological model building.

The paper is organized as follows. In Section~\ref{sec:ggnf}, we introduce the general tensorial formulation of the charge operator and derive its action on arbitrary tensor representations. In particular, we explicitly show how the resulting expression corresponds to the standard additivity of weights associated with Cartan generators. In Section~\ref{sec:applications}, we present several illustrative examples for different unitary simple or semi-simple groups. Finally, Section~\ref{sec:con} contains our conclusions and a brief discussion of possible extensions.

\section{Generalized Charge Tensor}
\label{sec:ggnf}

We aim to build the GMN formula to determine electric charges within multiplets of arbitrary dimensions. We will consider the operator $Q$ as a sum of all diagonal generators that are not necessarily related to the electric charge operator. In the fundamental representation, the operator $Q$ can be expressed as a straightforward extension of Eq.~\eqref{gnf}. In matrix form,
\begin{equation}
Q=\sum_j \alpha_jT_j+\frac{Y}{2}\mathbbm{1}
\label{gco}
\end{equation}
Here, $T_j$ represents the diagonal generators of the $SU(n)$ group, and $Y$ denotes the hypercharge or generator of the $U(1)_Y$ group. The coefficients $\alpha_j$ are constants chosen in a way that, together with $Y$, integer values for charges are obtained.

To illustrate this, let us examine two specific examples: $\Psi=n\otimes \overbar{n}$ and $\Phi=n\otimes n$ \cite{Castillo-Ruiz:2020kuw}. In this case, the operators $Q$ are defined by
\begin{equation}
Q\Psi=\sum_j \alpha_j[T_j,\Psi]+\frac{Y}{2}\Psi,\quad Q\Phi=\sum_j \alpha_j\{T_j,\Phi\}+\frac{Y}{2}\Phi,
\label{sggnf}
\end{equation}
where $[,]$ and $\{,\}$ represent the commutator and anticommutator, respectively. The commutator applies to fermions, while the anticommutator is used for bosons. 

The more general case can be obtained by direct products $n\otimes n\otimes\cdots $ or $n\otimes \overbar{n}\otimes\cdots$ of groups $SU(n)$, $SU(n)\otimes U(1)$, or $SU(n_1)\otimes\cdots SU(n_p)\otimes U(1)_1\otimes\cdots U(1)_q$. In these cases, the charge assignment for higher-dimensional representations cannot be arbitrary; it must be compatible with the definition of the charge operator of the respective model given in Eq.~(\ref{gco}). In this sense, we may call this method a generalized GMN formula if the latter is considered in the original $SU(2)\otimes U(1)$ case only.

The resulting charge values depend on the specific choices of $\alpha_j$ and $Y$. Consider, for example, the case of the electric charge in $SU(3)$ theory of mesons and baryons; each $SU(2)$ subgroup possesses a particular strong hypercharge that correlates with its associated flavor quantum number.

To begin explaining our approach, consider a general reducible tensor $t$ for the scenario $SU(n)\otimes U(1)_Y$, represented as $(t,Y)^{i_1i_2i_3\ldots i_p}_{j_1j_2j_3\ldots j_q}$. Here, the upper indices indicate that the tensor was built from the multiplication of fundamental representations $u^{i_1}u^{i_2}u^{i_3}\ldots u^{i_p}$, and the lower indices from the anti-fundamentals $u_{j_1}u_{j_2}u_{j_3}\ldots u_{j_q}$.

Lie groups $SU(n)\otimes U(1)_{Y}$ have elements that can be represented as $e^{-i\vv{\alpha}(x)\cdot \vv{T}}e^{-i\beta(x)\frac{Y}{2}}$, where $\vv{\alpha}(x)=\left(\alpha _{1}(x),\ldots,\alpha_{n^2-1}(x)\right)$ and $\beta(x)$ are the parameters of $SU(n)$ and $U(1)_Y$, respectively; while $\vv{T}=\left(T_{1},\ldots,T_{n^{2}-1}\right)$ and $Y$ are their generators. These group elements act on fields arranged into fundamental representations $q^i$, which we will denote in the usual dimensional notation: $q^i\sim (n,Y)^i$ with $i=1\ldots n$. The gauge transformation is
\begin{equation}
\left(n,Y\right)^{k}\rightarrow e^{-i\vv{\alpha}(x)\cdot \vv{T}}e^{-i\beta(x)\frac{Y}{2}} \left(n,Y\right)^{k}\approx \left(\mathbbm{1}_{n}-i\sum_{j=1}^{n^2-1}\alpha_j(x)\frac{\lambda_j}{2}-i\beta(x)\frac{Y}{2}\mathbbm{1}_n\right)\left(n,Y\right)^{k} .
\label{app1}
\end{equation}
We will denote the antifundamental representation by $q_k\sim (\overbar{n},\overbar{Y})_k$. Recall that for $SU(\bar{n})$, generators are $T_{\overbar{n}}^a=-\left(T_n^a\right)$, while for $U(1)$, charges change their sign $\overbar{Y}=-Y$.

Hence, the general charge operator for the (anti-)fundamental representation (of dimension $n$) in units of $\vert e\vert$ in the latter equation is
\begin{equation}\label{Q_n}
\begin{split}    
   Q(n,Y)^k&=\left(M^k_a + \frac{Y}{2}\delta_a^k\right)(n,Y)^a \\
    Q(\bar{n},\bar{Y})_k&=-\left(M^a_k + \frac{Y}{2}\delta^a_k\right)(\bar{n},\bar{Y})_a,
\end{split}
\end{equation}
where the $M^i_j$ tensors are linear combinations of the diagonal generators defined as
\begin{equation}\label{diag_gen}
M^k_a\equiv\sum_m\alpha_m\frac{(\lambda_m)^k_a}{2}.
\end{equation}

These equations are consistent with Eq.~\eqref{gco}, considering that $\beta$ is absorbed for the $Y$ value and index $m$ runs over the $n-1$ diagonal generators; in $SU(2):m=3$, in $SU(3):m=3,8$, and $SU(4)$ we have $m=3,8,15$, and so on. Observe that the considered charge operator is applied to the components of $(n,Y)$. This is because $Q$ operators live in $U(1)_Q$ symmetry, where the multiplets of $SU(n)\otimes U(1)_Y$ are decomposed into singlets.

Then, a multiplet with an arbitrary dimension is obtained from the product of any number of fundamental and anti-fundamental tensors and has the following representation:
\begin{equation}\label{T_tensor}
\left(t,Y\right)^{i_1i_2i_3\ldots i_p}_{j_1j_2j_3\ldots j_q}=\prod_{k,l=1}^{p,q}\left(n,Y_k\right)^{i_k}\left(\overbar{n},\bar{Y}_l\right)_{j_l}.
\end{equation}

Due to the additive nature of the abelian $U(1)_Y$ generators, the total hypercharge is $Y=\sum_{k=1}^pY_k+\sum_{l=1}^q\bar{Y}_l$. Additionally, this tensor is typically reducible, but the physically meaningful representations are irreducible and require  some derivation from conventional tensor decomposition. Suppose that there are $r_1$ fundamental and $r_2$   anti-fundamental irreducible representations obtained from the decomposition of $\left(t,Y\right)=(n,Y_k)\otimes(\bar{n},\bar{Y}_l)$; then Eq.~\eqref{T_tensor} could be written in terms of the irreps $(s,Y)$ as:
\begin{equation}\label{Tred_tensor}
    \left(t,Y\right)=\bigoplus_{k=1}^{r_1}\left(s_k,Y_k\right)\,\oplus\, \bigoplus_{l=1}^{r_2}\left(\bar{s}_l,\bar{Y}_l\right),
\end{equation}
where the tensor indices are omitted for the sake of simplicity. 

For example, in a theory with symmetry $SU(3)\otimes U(1)_Y$, using the last two equations, a second-order tensor is 
\begin{equation}\label{ex9}
(9,Y)^{i_1i_2}=(3,Y_1)^{i_1}(3,Y_2)^{i_2} = (6,Y_1)^{(i_1i_2)}\oplus(\overbar{3},Y_2)^{[i_1i_2]}.    
\end{equation}

Then, the transformation of the general tensor is 
\begin{equation} \label{T_gauge}
\begin{split}
 \left(t,Y\right)^{i_1i_2i_3\ldots i_p}_{j_1j_2j_3\ldots j_q}\rightarrow e^{i\beta(x)\frac{Y}{2}}&U^{i_1}_aU^{i_2}_b\ldots U^{i_p}_zU^{a'}_{j_1}U^{b'}_{j_2}\ldots U^{z'}_{j_q}\left(t,Y\right)^{ab\ldots z}_{a'b'\ldots z'}\\
    = e^{i\beta(x)\frac{Y}{2}}&\left[\bigoplus_{k=1}^{r_1}U^{i_1}_aU^{i_2}_b\ldots U^{i_p}_zU^{a'}_{j_1}U^{b'}_{j_2}\ldots U^{z'}_{j_q}\left(s_k,Y_k\right)^{ab\ldots z}_{a'b'\ldots z'}\right.\\
    &\left.\oplus\bigoplus_{l=1}^{r_2}U^{i_1}_aU^{i_2}_b\ldots U^{i_p}_zU^{a'}_{j_1}U^{b'}_{j_2}\ldots U^{z'}_{j_q}\left(\bar{s}_l,\bar{Y}_l\right)^{ab\ldots z}_{a'b'\ldots z'}\right].
\end{split}
\end{equation}

What follows is the construction of the charge operator for different multiplets in the respective representation, considering that the assignment of charges is meaningful only for physical fields (irreps) that result from the decompositions defined in Eq.~\eqref{Tred_tensor}. For that purpose, we need to express the transformations in Eq.~\eqref{T_gauge} as infinitesimal representations. By applying the operator defined in Eq.~\eqref{Q_n} as many times as the tensor $(t,Y)$ has indices, we obtain the electric charge operator for a general reducible tensor, whose components can be seen as  weight states of a special case of Cartan generators (see the Appendix).
\begin{equation}\label{A_formula}
\begin{split}
\left[Q(t,Y)\right]^{i_1i_2i_3\ldots i_p}_{j_1j_2j_3\ldots j_q}&\approx M^{i_1}_a(t,Y)^{ai_2\ldots i_p}_{j_1j_2\ldots j_q}+M^{i_2}_b(t,Y)^{i_1b\ldots i_p}_{j_1j_2\ldots j_q}+\ldots +M^{i_p}_z(t,Y)^{i_1i_2\ldots z}_{j_1j_2\ldots j_q}\\
&-M^{a'}_{j_1}(t,Y)^{i_1i_2\ldots i_p}_{a'j_2\ldots j_q}M^{b'}_{j_2}(t,Y)^{i_1i_2\ldots i_p}_{j_1b'\ldots j_q}-\ldots -M^{z'}_{j_1}(t,Y)^{i_1i_2\ldots i_p}_{j_1j_2\ldots z'}\\
&+\frac{Y}{2}(t,Y)^{i_1i_2i_3\ldots i_p}_{j_1j_2j_3\ldots j_q}.
\end{split}
\end{equation}

Equation \eqref{A_formula} provides an explicit index-level realization of the action of an abelian Cartan generator on tensor products, incorporating the standard additivity of weights along with the sign change associated with the dual representation. When irreducible representations are derived via Eq. \eqref{Tred_tensor}, electric charge eigenstates will be obtained. All entire process of assigning electric charges will depend on how the tensor decomposition of $(t,Y)$ is performed, as shown in the following section for some representative cases.

We can see that the $Q$ operator belongs to the Cartan subalgebra of the non-abelian group (plus a possible external $U(1)$) and acts diagonally in a weight basis, defining a hypercharge that satisfies $\comm{T}{Y}=0$.

In the fundamental representation, this operator is represented by the matrices $M=\sum_{i=1}^{N-1}\alpha_i\lambda_i$ in Eq. \eqref{diag_gen}, whose eigenvalues correspond to the weights of the fundamental multiplet projected along the direction defined by $Q$.

The generalized formula is nothing but the index-level realization of the additivity of weights in tensor products, together with the sign reversal associated with the dual representation. The present formulation provides a compact bookkeeping rule for computing charge eigenvalues directly at the tensor level, without the need to explicitly decompose the representation into irreducible components, as in Eq.~\eqref{ex9}.

In summary, the goal of this work is to obtain the $Q$ operator applied to all kinds of multiplets, demonstrating that it is independent of whether any multiplet couples with other known particles or not. We only use the properties of the Lie groups to which the multiplets belong. This method could be used for dark-matter multiplets such as $SU(2)$ vector doublets \cite{PhysRevD.99.075026}, general fermionic n-tuplets of $SU(2)$ \cite{CIRELLI2006178}, general scalar n-tuplets in the symmetry of the SM \cite{Beauchesne2024}, and so on.

\section{Applications}
\label{sec:applications}

In this section, we apply the method discussed above by considering representations in the following cases: A) $SU(2)\otimes U(1)$ as in the strong isospin formalism or in the electroweak standard model; B) $SU(3)\otimes U(1)$ with a general $\alpha_8$ that defines these types of theories; C) $SU(5)$, which contains a color symmetry that has nothing to do with electric charges, and D) special cases including the color $SU(3)_C$ and horizontal symmetries. The following examples do not propose any new rules for charge quantization; rather, they demonstrate how the tensor-index formulation can be used in practice as a bookkeeping device for already known representations.

\subsection{\texorpdfstring{$SU(2)\otimes U(1)_Y$}{SU2U1}}
\label{subsec:sml}

Let us begin with the simplest case to illustrate the method. In this symmetry, equation \eqref{diag_gen} is
\begin{equation}
M_{3}=\alpha_3\frac{\tau_3}{2}=\frac{1}{2}\begin{pmatrix}\alpha_3 & 0 \\ 0 & -\alpha_3\end{pmatrix}. \end{equation}

Firstly, we consider the fundamental representation with only one upper index $(2,Y)^i$, the simplest case, to clarify our notation. The formula \eqref{A_formula} for this case is
\begin{equation}\label{SM_fund_form}
\left[Q(2,Y)\right]^i=\left[M_{3}\right]^i_a(2,Y)^a+\frac{Y}{2}(2,Y)^i,
\end{equation}
with $a=1,2$. Explicitly, the components of $M_{3}$ are $\left[M_{3}\right]^1_1=\frac{1}{2}\alpha_3$, $\left[M_{3}\right]^2_2=-\frac{1}{2}\alpha_3$, and $\left[M_{3}\right]^1_2=\left[M_{3}\right]^2_1=0$. Then expanding \eqref{SM_fund_form},
\begin{equation}
\begin{split}
    \left[Q(2,Y)\right]^1&=\left[M_{3}\right]^1_1(2,Y)^1+\frac{Y}{2}(2,Y)^1=\frac{1}{2}\left(\alpha_3+Y\right)(2,Y)^1\\
    \left[Q(2,Y)\right]^2&=\left[M_{3}\right]^2_2(2,Y)^2+\frac{Y}{2}(2,Y)^2=\frac{1}{2}\left(-\alpha_3+Y\right)(2,Y)^2
\end{split}
\end{equation}

These eigenvalues coincide with the diagonal values of Eq.~\eqref{Q_n}. We represent \eqref{SM_fund_form} in the form of an $i$-ordered matrix,
\begin{equation}\label{Q_fund}
    Q(2,Y)\rightarrow\frac{1}{2}\begin{pmatrix} \alpha_3+Y \\ -\alpha_3+Y \end{pmatrix}
\end{equation}

For antifundamental representations, a lower index is used in Eq. \eqref{A_formula}, and the electric charges have the opposite sign to those in Eq. \eqref{Q_fund}. 

In the Electroweak Standard Model (ESM) with $\alpha_3=1$, it is usual to set $Y=-1$ for the left leptonic doublet, $Y=1$ for the scalar case, and $Y=1/3$ for the quark doublets. These choices maintain the usual fields in the correct order in the doublets.

The case of $\mathbf{2}\otimes\mathbf{2}$ is represented by tensors with two upper indices $t^{i_ii_2}$. The generalized formula \eqref{A_formula} is
\begin{equation}
\left[Q(t,Y)\right]^{i_1i_2}=\left[M_{\alpha}\right]^{i_1}_a\left[(t,Y)\right]^{ai_2}+\left[M_{\alpha}\right]^{i_2}_b\left[(t,Y)\right]^{i_1b}+\frac{Y}{2}\left[(t,Y)\right]^{i_1i_2}.
\end{equation}

The irreducible representations are obtained from $t^{i_1i_2}=u^{i_1}u^{i_2}=u^{(i_1}u^{i_2)}+u^{[i_1}u^{i_2]}$ or $\mathbf{2}\otimes \mathbf{2} = \mathbf{3_S} \oplus \mathbf{1_A}$. Due to the linearity of the charge operator, $Q\left(u^{(i_1}u^{i_2)}+u^{[i_1}u^{i_2]}\right)=Q(3_S,Y)^{(i_1i_2)}+Q(1_A,Y)^{[i_1i_2]}$.
\begin{equation}\label{SU2_trip_sing}
\begin{split}
\left[Q(3_S,Y)\right]^{(i_1i_2)}&=\left[M_{\alpha}\right]^{i_1}_a\left[(t,Y)\right]^{(ai_2)}+\left[M_{\alpha}\right]^{i_2}_b\left[(t,Y)\right]^{(i_1b)}+\frac{Y}{2}\left[(t,Y)\right]^{(i_1i_2)},\\
\left[Q(1_A,Y)\right]^{[i_1i_2]}&=\left[M_{\alpha}\right]^{i_1}_a\left[(t,Y)\right]^{[ai_2]}+\left[M_{\alpha}\right]^{i_2}_b\left[(t,Y)\right]^{[i_1b]}+\frac{Y}{2}\left[(t,Y)\right]^{[i_1i_2]}.
\end{split}
\end{equation}

After performing the calculations in the former case, there are three eigenvalues, while in the latter case, there exists only one for $i_1,i_2=1,2$.
\begin{equation}\label{Q_SM_3_1}
 Q(3_S,Y)=\frac{1}{2}\begin{pmatrix} 2\alpha_3+Y \\ Y \\ -2\alpha_3+ Y \end{pmatrix},\quad Q(1_A,Y)=\frac{1}{2} Y.
\end{equation}

As usual, we set $\alpha_3=1$, consistent with nearly all theories. For example, in SM, the vector bosons ($Y=0$) have eigenvalues $\begin{pmatrix}+1 & 0 & -1\end{pmatrix}^T$; this is consistent with the usual commutator for acquiring charges when the triplet is in its adjoint representation. Another example is the type of multiplets theorized in diquark models \cite{PhysRevLett.48.1653}, theoretical states that would be arranged in triplets with $Y=2/3$ to obtain $\begin{pmatrix}4/3 & 1/3 & -2/3\end{pmatrix}^T$ electric charges.

In the same symmetry group, there is another possibility: a mixed tensor $t_i^j=u_iu^j$ or $\overbar{\mathbf{2}}\otimes\mathbf{2}$. In the same way as in the previous case, irreducible representations are known by decomposing $t_j^i=u^iu_j=\left(u^iu_j-\frac{1}{2}\delta_j^i u^ku_k\right)+\left(\frac{1}{2}\delta_j^i u^ku_k\right)$ or $\overbar{\mathbf{2}}\otimes \mathbf{2}=\mathbf{3} \oplus \mathbf{1_S}$. Then 
\begin{equation}\label{SM_trip}
\left[Q(t,Y)\right]^i_j=\left[M_{\alpha}\right]^i_a\left[(t,Y)\right]^a_j-\left[M_{\alpha}\right]_j^b\left[(t,Y)\right]^i_b+\frac{Y}{2}\left[(t,Y)\right]^i_j.
\end{equation}

The eigenvalues are
\begin{equation}
 Q(3,Y)=\frac{1}{2}\begin{pmatrix} 2\alpha_3+Y  \\ Y \\ -2\alpha_3+ Y \end{pmatrix},\quad Q(1_S,Y)=\frac{1}{2} Y.
\end{equation}

Pions and rho mesons fall into this class of triplets, as their states are made up of a quark and an antiquark doublet from the first generation. With $\alpha_3=1$ and $Y=0$, the electric charges are $\begin{pmatrix}+1 & 0 & -1\end{pmatrix}^T$. The order in which the charges are written does not matter, as they are obtained individually according to \eqref{SM_trip}.

Another test to perform is for the tensor $t^{i_1i_2i_3}=u^{i_1}u^{i_2}u^{i_3}=u^{(i_1}u^{i_2}u^{i_3)}$ $+\frac{1}{3}\epsilon_{i_1i_3}u^{(i_1}u^{i_2)}u^{i_3}+\frac{1}{3}\epsilon_{i_2i_3}u^{(i_1}u^{i_2)}u^{i_3}$ or $\mathbf{2}\otimes \mathbf{2}\otimes \mathbf{2}\equiv\mathbf{2^{\otimes 3}}=\mathbf{4_S}\oplus 2(\mathbf{2})$. In this case, Eq.~\eqref{A_formula} is
\begin{equation}\label{C_tensor}
\begin{split}
\left[Q(4_S,Y)\right]^{(i_1i_2i_3)}&=\left(M_{\alpha}\right)^{i_1}_a\left[(4_S,Y)\right]^{(ai_2i_3)}+\left(M_{\alpha}\right)^{i_2}_b\left[(4_S,Y)\right]^{(i_1bi_3)}+\\ &+\left(M_{\alpha}\right)^{i_3}_c \left[(4_S,Y)\right]^{(i_1i_2c)}+\frac{Y}{2}\left[(4_S,Y)\right]^{(i_1i_2i_3)}.    
\end{split}
\end{equation}

The $Q$ eigenvalues for the quartet (the doublets have already been discussed) are
\begin{equation}
    Q(4_S,Y)=\frac{1}{2}\begin{pmatrix}3\alpha_3+ Y \\ \alpha_3+Y \\ -\alpha_3+Y \\ -3\alpha_3+Y \end{pmatrix}
\end{equation}

Provided that $\alpha_3=1$ and $Y=1$, the resulting electric charges $\begin{pmatrix}2 & 1 & 0 & -1\end{pmatrix}^T$ are aligned with the Delta quartet featured in the eight-fold theory~\cite{Gell-Mann:1956iqa}.

If we continue with the same procedure, we can find the electric charges for the quintet, sextet, etc. These are the totally symmetric addends of the fundamental product decomposition in four dimensions, five dimensions, and so on. We mean $\mathbf{2^{\otimes 4}}=\mathbf{5_S}\oplus (3)\mathbf{3} \oplus (2)\mathbf{1}$, $\mathbf{2^{\otimes5}} =\mathbf{6_S}\oplus (4)\mathbf{4} \oplus (2)\mathbf{5}$, or $\mathbf{2^{\otimes 6}} =\mathbf{7_S}\oplus (5)\mathbf{5} \oplus (9)\mathbf{3}\oplus (5)\mathbf{1}$. Then
\begin{equation}
Q(5_S,Y)=\begin{pmatrix}2\alpha_3+\frac{Y}{2} \\ \alpha_3+\frac{Y}{2} \\ \frac{Y}{2} \\ -\alpha_3+\frac{Y}{2} \\ -2\alpha_3+\frac{Y}{2} \end{pmatrix},\; Q(6_S,Y)=\frac{1}{2}\begin{pmatrix}5\alpha_3+Y \\ 3\alpha_3+Y \\ \alpha_3+Y \\ -\alpha_3+Y \\ -3\alpha_3+Y \\ -5\alpha_3+Y\end{pmatrix},\; Q(7_S,Y)=\begin{pmatrix}3\alpha_3+\frac{Y}{2} \\ 2\alpha_3+\frac{Y}{2} \\ \alpha_3+\frac{Y}{2} \\ \frac{Y}{2} \\ -\alpha_3+\frac{Y}{2} \\ -2\alpha_3+\frac{Y}{2} \\ -3\alpha_3+\frac{Y}{2}\end{pmatrix}.
\end{equation}

This opens up the opportunity to study the electric charges of tetraquarks \cite{PhysRevLett.118.022003} represented as quintets, triplets, or singlets, where we must include zero-charged multiplets of the form $qq\bar{q}\bar{q}\sim u^{i_1}u^{i_2}u_{j_1}u_{j_2}$, choosing the adequate hypercharge. The same applies to pentaquarks \cite{PhysRevLett.115.072001} represented as quartets, quintets, or sextets; or even the theorized hexaquarks \cite{PhysRevD.85.014019}.

\subsection{\texorpdfstring{$SU(3)\otimes U(1)_Y$}{SU3U1}}
\label{subsec:331l}

This case occurs in the eight-fold way \cite{Gell-Mann:1956iqa} and in some electroweak models based on symmetry $SU(3)_L\otimes U(1)_Y$, like the ones described in \cite{Singer:1980sw,Pisano:1991ee,Foot:1992rh,Frampton:1992wt}. The matrix \eqref{diag_gen} in this group is
\begin{equation}
    M_{\alpha}=\frac{\alpha_3}{2}\lambda_3+\frac{\alpha_8}{2}\lambda_8= \frac{1}{2}\begin{pmatrix}\alpha_3+\frac{\alpha_8}{\sqrt{3}} & 0 & 0 \\ 
    0 & -\alpha_3+\frac{\alpha_8}{\sqrt{3}} & 0 \\ 0 & 0 & -\frac{2\alpha_8}{\sqrt{3}} \end{pmatrix}
\end{equation}
The eigenvalues of the fundamental charge operator (one index only) are:
\begin{equation}\label{Charge_op_331}
    \left[Q(3,Y)\right]^i=\left[M_{\alpha}\right]^i_a \left[(3,Y)\right]^a+\frac{Y}{2}\left[(3,Y)\right]^i\rightarrow 
    \begin{pmatrix} \frac{\alpha_3}{2}+\frac{\alpha_8}{2\sqrt{3}}+ \frac{Y}{2} \\ -\frac{\alpha_3}{2}+\frac{\alpha_8}{2\sqrt{3}}+ \frac{Y}{2} \\ -\frac{\alpha_8}{\sqrt{3}} + \frac{Y}{2}\end{pmatrix}.
\end{equation}

A prospect used in any phenomenological treatment \cite{Singer:1980sw,Pisano:1991ee,Foot:1992rh,Frampton:1992wt} to obtain charge quantization is $\alpha_3=1,\,\alpha_8=\pm\sqrt{3},\pm\frac{1}{\sqrt{3}}$ (in the literature $\alpha_8\rightarrow \beta$), where SM doublets are recovered. Subsequently, the fermion (or scalar) triplet may obtain different electric charge values, depending on the $\alpha_8$' choice.
\begin{center}
\begin{tabular}{ r c c c c }
 $\alpha_8=$ & $\sqrt{3}$ & $-\sqrt{3}$ & $\frac{1}{\sqrt{3}}$ & $-\frac{1}{\sqrt{3}}$\\ 
 $Q\left(3,Y\right)=$ & $\begin{pmatrix}1+\frac{Y}{2} \\ \frac{Y}{2} \\ -1+\frac{Y}{2} \end{pmatrix},$ & $\begin{pmatrix}\frac{Y}{2} \\ -1+\frac{Y}{2} \\ 1+\frac{Y}{2} \end{pmatrix},$ & $\begin{pmatrix}\frac{2}{3}+\frac{Y}{2} \\ -\frac{1}{3}+\frac{Y}{2} \\ -\frac{1}{3}+\frac{Y}{2} \end{pmatrix},$ & $\begin{pmatrix}\frac{1}{3}+\frac{Y}{2} \\ -\frac{2}{3}+\frac{Y}{2} \\ \frac{1}{3}+\frac{Y}{2} \end{pmatrix}.$\\
\end{tabular}
\end{center}

If we want to recover the fermionic (or scalar) doublets of the SM located in the two upper positions of the triplets, the hypercharge has to be set to $Y=-2,\,0,\,-4/3 \text{ or } -2/3$, respectively, and the third component would be the electric charge of a new exotic fermion like the heavy $E^+$ if $\alpha_8=-\sqrt{3}$ is chosen \cite{PhysRevD.48.2353}. The same reasoning is used to recover the scalar doublet of the SM. The choice of $\alpha_8$ changes the particle content and the whole theory.

Continuing with the testing, as in the SM symmetry, there are three different cases when considering two-index representations:
\begin{align}
(i)\enspace & 3\otimes 3 = 6_S \oplus \overbar{3}_A: t^{i_1 i_2}=u^{i_1}u^{i_2}=u^{(i_1}u^{i_2)}+ u^{[i_1}u^{i_2]} \nonumber\\ 
&\left[Q(6_S,Y)\right]^{(i_1i_2)}=\left[M_{\alpha}\right]^{i_1}_a\left[(6_S,Y)\right]^{(ai_2)}+\left[M_{\alpha}\right]^{i_2}_b\left[(6_S,Y)\right]^{(i_1b)}+\frac{Y}{2}\left[(6_S,Y)\right]^{(i_1i_2)}\\   \nonumber
& \xrightarrow[]{\alpha_3=1}\;\begin{pmatrix} 1+\frac{\alpha_8}{\sqrt{3}}+\frac{Y}{2} & \frac{\alpha_8}{\sqrt{3}}+\frac{Y}{2} & \frac{1}{2}-\frac{\alpha_8}{2\sqrt{3}}+\frac{Y}{2} \\ & -1+\frac{\alpha_8}{\sqrt{3}}+\frac{Y}{2} & -\frac{1}{2}-\frac{\alpha_8}{2\sqrt{3}}+\frac{Y}{2} \\ & &  -2\frac{\alpha_8}{\sqrt{3}}+\frac{Y}{2} \end{pmatrix}. \\
(ii)\enspace & \overbar{3}\otimes \overbar{3} = \overbar{6}_S \oplus 3_A, \text{the anti-sextet has the opposite charges of $(6_S,Y)$.}\\ \nonumber
(iii)\enspace & 3\otimes \overbar{3}= 8 \oplus 1: t^i_j=u^iu_j=\left(u^iu_j-\frac{1}{3}\delta_j^i u^ku_k\right)+\left(\frac{1}{3}\delta_j^i u^ku_k\right)\\ 
&\left[Q(8,Y)\right]^i_j=\left[M_{\alpha}\right]^{i}_a\left[(8,Y)\right]^{a}_j- \left[M_{\alpha}\right]^{b}_j\left[(8,Y)\right]^{i}_b+\frac{Y}{2}\left[(8,Y)\right]^i_j \nonumber\\
&\xrightarrow[]{\alpha_3=1}\;\dfrac{1}{2}\begin{pmatrix} Y & 2+ Y & 1+\sqrt{3}\alpha_8+ Y \\ -2+ Y & Y & -1+\sqrt{3}\alpha_8+ Y \\ -1-\sqrt{3}\alpha_8+ Y & 1-\sqrt{3}\alpha_8+ Y & Y
\end{pmatrix}.\label{31octet}
\end{align}

The most widely used sextet is the one defined with $\alpha_8=-\sqrt{3}$ and $Y=0$ \cite{Foot:1992rh}, which is required to complete the masses for all leptons and to include Majorana mass terms for neutrinos, allowing the Seesaw mechanism. However, there are other possibilities,
\begin{center}
\begin{tabular}{ r c c }
 $\alpha_8=$ & $\sqrt{3}$ & $-\sqrt{3}$\\ 
 $Q\left(6_S,Y\right)=$ & $\begin{pmatrix} 2+\frac{Y}{2} & 1+\frac{Y}{2} & \frac{Y}{2} \\ & \frac{Y}{2} & -1+\frac{Y}{2} \\ & &  -2+\frac{Y}{2} \end{pmatrix},$ & $\begin{pmatrix} \frac{Y}{2} & -1+\frac{Y}{2} & 1+\frac{Y}{2} \\  & -2+\frac{Y}{2} & \frac{Y}{2} \\  &  & 2+\frac{Y}{2} \end{pmatrix},$\\ \\
$\alpha_8=$ & $\frac{1}{\sqrt{3}}$ & $-\frac{1}{\sqrt{3}}$ \\ 
$Q\left(6_S,Y\right)=$ & $\begin{pmatrix} \frac{4}{3}+\frac{Y}{2} & \frac{1}{3}+\frac{Y}{2} & \frac{1}{3}+\frac{Y}{2} \\  & -\frac{2}{3}+\frac{Y}{2} & -\frac{2}{3}+\frac{Y}{2} \\  &  & -\frac{2}{3}+\frac{Y}{2} \end{pmatrix},$ & $\begin{pmatrix} \frac{2}{3}+\frac{Y}{2} & -\frac{1}{3}+\frac{Y}{2} & \frac{2}{3}+\frac{Y}{2} \\  & -\frac{4}{3}+\frac{Y}{2} & -\frac{1}{3}+\frac{Y}{2} \\  &  & \frac{2}{3}+\frac{Y}{2} \end{pmatrix}$
\label{331sextetos}
\end{tabular}
\end{center}

Otherwise, a general sextet can be formed by scalar, fermion, or vector particles, and all of them may have an arbitrary electric charge depending on the value of $Y$. For example, if $\abs{Y}=6$ all components are charged for any $\alpha_8$. Of course, not all attributions will be phenomenologically safe. For scalars, these kinds of sextets (and all the others with $\abs{Y}>6$) do not perform any vev, nor do they couple with the usual fermions; however, they could couple with the vector gauge bosons of the model.

For the octet case, there is an open possibility of considering them within $SU(3)_L\otimes U(1)_Y$ theories. For example, adding three generations of leptonic octet representations $\sim(8,0)$ can provide gauge coupling unification for a TeV breaking generation \cite{DEPPISCH2016432}, while driving an interesting radiative model for neutrino mass generation \cite{PhysRevD.91.031702}. 
The possible octets are
\begin{center}
\begin{tabular}{ r c c c c}
 $\alpha_8=$ & $\sqrt{3}$ & $-\sqrt{3}$ \\ 
 $Q\left(8,Y\right)=$ & $\dfrac{1}{2}\begin{pmatrix} Y & 2+ Y & 4+ Y \\ -2+ Y & Y & 2+Y \\ -4+ Y & -2+ Y & Y\end{pmatrix},$ & $\dfrac{1}{2}\begin{pmatrix} Y & 2+ Y & -2+ Y \\ -2+ Y & Y & -4+ Y \\ -2+ Y & 4+ Y & Y\end{pmatrix}$, \\
 \\
 $\alpha_8=$ &  $\frac{1}{\sqrt{3}}$ & $-\frac{1}{\sqrt{3}}$ \\
  & $\dfrac{1}{2}\begin{pmatrix} Y & 2+Y & 2+Y \\ -2+Y & Y & Y \\ -2+Y & Y & Y\end{pmatrix},$ & $\dfrac{1}{2}\begin{pmatrix} Y & 2+Y & Y \\ -2+Y & Y & -2+Y \\ Y & 2+Y & Y \end{pmatrix}$
\end{tabular}
\end{center}

It is interesting to analyze when the electric charges align with the quark configurations in the meson octet. In the $SU(3)_C$ symmetry, there is no breaking, and only the strong force is considered, allowing us to set $Y=0$. The electric charges in the meson octet align when $\alpha_8=\frac{1}{\sqrt{3}}$. 

There exists an opportunity to explore other types of octets in a gauge model incorporating exotic vectorial leptons analogous to the octets described in \cite{PhysRevD.71.015009}, as detailed in Eq. \eqref{31octet} for any $\alpha_8$.

Finally, in $SU(3)\otimes U(1)$ symmetry as well, we analyze $3^{\otimes 3}=10_S \oplus 2(8) \oplus 1$. The charge operator for the decuplet is a tensor with three upper indices that is totally symmetric in $i_ii_2i_3$.
Then, if $\alpha_3=1$,
\begin{equation}
\begin{split}
    \left[Q(10,Y)\right]^{(i_ii_2i_3)} &= \left(M_{\alpha}\right)^{i_1}_a\left[(10,Y)\right]^{(ai_2i_3)} + \left(M_{\alpha}\right)^{i_2}_b\left[(10,Y)\right]^{(i_1bi_3)}\\
    & + \left(M_{\alpha}\right)^{i_3}_c\left[(10,Y)\right]^{(i_1i_2c)} +\frac{Y}{2}\left[(10,Y)\right]^{(i_1i_2i_3)}
\end{split}
\end{equation}

\begin{equation}\label{qdecuplet}
Q(10,Y)\rightarrow \frac{1}{2}\begin{pmatrix}3+\sqrt{3}\alpha_8 +Y \\ 1+\sqrt{3}\alpha_8+ Y \\ -1+\sqrt{3}\alpha_8+ Y \\ -3+\sqrt{3}\alpha_8+Y  \\ 1-\sqrt{3}\alpha_8 +Y\\ -1-\sqrt{3}\alpha_8+Y\\ -2\sqrt{3}\alpha_8 +Y \\ 2+Y \\ Y \\ -2+Y \end{pmatrix}
\end{equation}

Similarly to the octet case, setting $Y=0$ and selecting $\alpha_8 = \frac{1}{\sqrt{3}}$ yield the observed baryon decuplet. For particles with non-zero charge, setting $Y=4$ results in exotic electric charges of $+4$ and $+3$. Moreover, certain experimental findings, such as the high-precision jet data from the ATLAS experiment, necessitate these types of decoupling for exotic fermions \cite{LLORENTE2018106}. 

If the decuplet $\Delta$ were a scalar, it would be coupled with scalar triplets $\eta$, allowing terms such as $\eta\eta\eta\Delta^*$ in the scalar potential.

\subsection{\texorpdfstring{$SU(5)$}{SU5}}
\label{sec:su5}

In the $SU(5)$ grand unified theory \cite{Georgi:1974sy}, the quarks and leptons are arranged in the same fundamental representation, whose electric charges we want to reproduce as a final test of the method presented in this work. For this purpose, the hypercharges of the first three components differ from those of the last two.

The four diagonal generators are

\begin{equation}
    \begin{split}
        \lambda_3&=\diag\left(1,-1,0,0,0\right)\\
        \lambda_8&=\frac{1}{\sqrt{3}}\diag\left(1,1,-2,0,0\right)\\
        \lambda_{15}&=\diag\left(0,0,0,1,-1\right)\\
        \lambda_{24}&=\frac{1}{\sqrt{15}}\diag\left(-2,-2,-2,3,3\right)
    \end{split}
\end{equation}

Then, the matrix \eqref{diag_gen} is
\begin{equation}
M_{\alpha}=\frac{1}{2}
\begin{pmatrix}
\alpha_3+\frac{\alpha_8}{\sqrt{3}}-\frac{2\alpha_{24}}{\sqrt{15}} & 0 & 0 & 0 & 0 \\
0 & -\alpha_3+\frac{\alpha_8}{\sqrt{3}}-\frac{2\alpha_{24}}{\sqrt{15}} & 0 & 0 & 0\\
0 & 0 & -\frac{2\alpha_8}{\sqrt{3}}-\frac{2\alpha_{24}}{\sqrt{15}} & 0 & 0 \\
0 & 0 & 0 & \alpha_{15}+\frac{3\alpha_{24}}{\sqrt{15}} & 0 \\
0 & 0 & 0 & 0 & -\alpha_{15}+\frac{3\alpha_{24}}{\sqrt{15}}
\end{pmatrix}    
\end{equation}

The tensor \eqref{A_formula} applied to the one-index representation (quintet) is 
\begin{equation}\label{Charge_op_SU5}
\begin{split}
    \left[Q(5,Y_1,Y_2)\right]^i&=\left[M_{\alpha}\right]^i_a \left[(5,Y_1,Y_2)\right]^a+\frac{Y_{1,2}}{2}\left[(5,Y_1,Y_2)\right]^i \\
    &\rightarrow 
    \frac{1}{2}\begin{pmatrix} \alpha_3+\frac{\alpha_8}{\sqrt{3}}-\frac{2\alpha_{24}}{\sqrt{15}}+ Y_1 \\ -\alpha_3+\frac{\alpha_8}{\sqrt{3}}-\frac{2\alpha_{24}}{\sqrt{15}} + Y_1 \\ -\frac{2\alpha_8}{\sqrt{3}}-\frac{2\alpha_{24}}{\sqrt{15}} + Y_1 \\ \alpha_{15}+\frac{3\alpha_{24}}{\sqrt{15}} + Y_2 \\ -\alpha_{15}+\frac{3\alpha_{24}}{\sqrt{15}} + Y_2 \end{pmatrix}.
\end{split}
\end{equation}

In the usual representation \cite{Langacker:1980js}, the lepton doublet $Q\begin{pmatrix}e^+ & -\nu_e^c\end{pmatrix}^T\rightarrow \begin{pmatrix}+1 & 0\end{pmatrix}^T$ is written in the lower part of the quintet with hypercharge $Y_2$. Then, according to \eqref{Q_fund}, we assign $\alpha_3=1\Rightarrow Y_2=+1$, and
\begin{equation*}
    \frac{1}{2}\begin{pmatrix} \alpha_{15}+\frac{3\alpha_{24}}{\sqrt{15}} + 1 \\ -\alpha_{15}+\frac{3\alpha_{24}}{\sqrt{15}} + 1\end{pmatrix} = \begin{pmatrix} + 1 \\ 0 \end{pmatrix}\Rightarrow \alpha_{15}=1\,,\alpha_{24}=0.
\end{equation*}

With these constant values, the upper triplet of the quintet is the fundamental representation of $331$ theories, as seen in \eqref{Charge_op_331} if $Y_1=Y$. For example, with $\alpha_8=\frac{1}{\sqrt{8}},\,\alpha_{15}=1$ and $\alpha_{24}=0$, we have to set $Y_1=0$ to obtain the quark triplet.
\begin{equation}
Q\begin{pmatrix}u & d & q\end{pmatrix}^T\rightarrow \begin{pmatrix}2/3 & -1/3 & -1/3\end{pmatrix}^T.
\end{equation}

All these calculations are assumed to yield a triplet of flavor symmetry $SU(3)$ in $SU(3)\otimes SU(2)\otimes U(1)$. However, the GUT symmetry $SU(5)$ was thought to include the SM color $SU(3)_C$ with the decomposition $\textbf{5}=(3,1,Y_1)\oplus (1,2,Y_2)$. The electric charge of the upper triplet will depend on the particle content we choose.

The usual election \cite{Langacker:1980js,Georgi:1974sy} is $\begin{pmatrix}d^1 & d^2 & d^3 & e^+ & -\nu_e\end{pmatrix}_L^T\sim (3,1,Y_1)\oplus (1,2,Y_2)$, and we have to inspect the related hypercharges in the SM context. There, the electric charges of the quark singlet and lepton doublet via \eqref{Q_fund} and \eqref{Q_SM_3_1} are

\begin{equation}
    Q_q\left(3,1,-2/3\right)= -1/3,\;Q_l\left(1,2,1\right)=\begin{pmatrix}-1 \\ 0 \end{pmatrix}
\end{equation}

Then, decomposing $\textbf{5}=(3,1,-2/3)\oplus (1,2,1)$, the electric charge eigenvalues for the quintet are
\begin{equation}
    Q\left[(3,1,-2/3)\oplus (1,2,1)\right]=\begin{pmatrix}-1/3 \\ -1/3 \\ -1/3\end{pmatrix}\oplus\begin{pmatrix}-1 \\ 0 \end{pmatrix}=\begin{pmatrix}-1/3 \\ -1/3 \\ -1/3 \\ -1 \\ 0\end{pmatrix}.
\end{equation}

The other multiplets are built from $\mathbf{5}\otimes\mathbf{5}=\mathbf{10}_A\oplus \mathbf{15}_S$ and are represented by tensors with two upper indices. The irreducible representations are $t^{i_1i_2}=u^{i_1}u^{i_2}=u^{(i_1}u^{i_2)}+u^{[i_1}u^{i_2]}$.
\begin{equation}
\begin{split}
\left[Q(10_A,Y)\right]^{i_1i_2}&=\left[M_{\alpha}\right]^{i_1}_a\left[(t,Y)\right]^{[ai_2]}+\left[M_{\alpha}\right]^{i_2}_b\left[(t,Y)\right]^{[i_1b]}+\beta\frac{Y}{2}\left[(t,Y)\right]^{[i_1i_2]},\\
\left[Q(15_S,Y)\right]^{i_1i_2}&=\left[M_{\alpha}\right]^{i_1}_a\left[(t,Y)\right]^{(ai_2)}+\left[M_{\alpha}\right]^{i_2}_b\left[(t,Y)\right]^{(i_1b)}+\beta\frac{Y}{2}\left[(t,Y)\right]^{(i_1i_2)}.
\end{split}
\end{equation}

It is possible to compute the electric charges for these multiplets if we would like to work in $SU(5)$ theories with flavor symmetries only. However, we wish to find these charges for the actual $SU(5)$, with just one flavor in the upper triplet of the quintet. 

For that purpose, the decomposition $\textbf{5}\otimes \textbf{5}=\textbf{10}\oplus\textbf{15}$ can be decomposed into the SM subalgebra.

\begin{equation}
\begin{split}
\left[(3,1,-2/3)\oplus (1,2,1)\right]\otimes \left[(3,1,-2/3)\oplus (1,2,1)\right]&=\left[(1,1,2)\oplus (\bar{3},1,-4/3)\oplus (3,2,1/3)\right]\\ \oplus &\left[(1,3,2)\oplus (3,2,1/3)\oplus (6,1,-4/3)\right]
\end{split}
\end{equation}

Then
\begin{equation}\label{Q_SU5_dec}
\begin{split}
    Q(\textbf{10})&=Q(1,1,2)\oplus Q(\bar{3},1,-4/3)\oplus Q(3,2,1/3)=1\oplus\begin{pmatrix}-2/3 \end{pmatrix}_a \oplus \begin{pmatrix}(2/3)_a \\ (-1/3)_a \end{pmatrix} \\
    Q(\textbf{15})&=Q(1,3,2)\oplus Q(3,2,1/3)\oplus Q(6,1,-4/3)=\begin{pmatrix}0 \\ 1 \\ 2 \end{pmatrix} \oplus \begin{pmatrix} (2/3)_a \\ (-1/3)_a \end{pmatrix}\oplus\begin{pmatrix}-2/3\end{pmatrix}_{ab} \\
\end{split}
\end{equation}
where $a,b=1,2,3$ are color indices.

The decuplet $10_A$ is arranged in an antisymmetric matrix of fermions and is usually represented as \cite{Langacker:1980js}

\begin{equation}
\textbf{10}=\begin{pmatrix}
0 & u_3^c & -u_2^c & -u_1 & -d_1 \\ -u_3^c & 0 & u_1^c & -u_2 & -d_2 \\ u_2 & -u_1^c & 0 & -u_3 & -d_3 \\ u_1 & u_2 & u_3 & 0 & -e^+ \\ d_1 & d_2 & d_3 & e^+ & 0
\end{pmatrix}_L \rightarrow Q(10)=\begin{pmatrix}
 &  & \color{red} -2/3 & \color{red}-2/3 & \color{blue}+2/3 & \color{blue}-1/3 \\ &  &  & \color{red}-2/3 & \color{blue}+2/3 & \color{blue}-1/3 \\ &  &  &  & \color{blue}+2/3 & \color{blue}-1/3 \\  & &  &  &  & +1 \\  &  &  &  & 
\end{pmatrix},
\end{equation}
which can be built up from $\textbf{10}\rightarrow q^{[i_1}q^{i_2]}$, being $\textbf{5}\rightarrow q^i$. The electric charges in \eqref{Q_SU5_dec} can be verified in the last matrix representation.

\subsection{Other Symmetries}
\label{otsym}
In the $SU(3)_C$ symmetry, we have the fundamental representation $q^a=(q^r \,q^g\,q^b)^T$ and the anti-fundamental representation $\bar{q}=(q^{\bar{r}}\,q^{\bar{g} }\,q^{\bar{b}})^T$. 

Using Eq.~(27) with $\alpha_8=-\sqrt{3}$ and $Y=0$, we obtain the colored symmetric sextet
\begin{equation}
Q(\textbf{6})=\begin{pmatrix}gg & gb & gr\\ & bb & br \\  &  & rr \end{pmatrix}.
\label{colosextet}
\end{equation} 

In this kind of symmetry, the charges are not numbers but letters $r, g, b$ as usual. However, we could assign numbers, assuming they are summative quantities for each composite field. The same is done for the octet, demonstrating that the method is valid for any conserved charge within the framework of gauge theories. For the octet, we can use \eqref{31octet}
\begin{equation}
Q(\textbf{8})=\begin{pmatrix}\frac{2r\bar{r}-g\bar{g}-b\bar{b}}{3} & r\bar{g} & b\bar{g}\\ g\bar{r} & \frac{2g\bar{g}-r\bar{r}-b\bar{b}}{3} & b\bar{r} \\ g\bar{b} & r\bar{b} & \frac{2b\bar{b}-r\bar{r}-g\bar{g}}{3} \end{pmatrix}.
\label{coloroctet}
\end{equation}

It must be recalled that it is an octet because the last state on the diagonal is regarded as a linear combination of the other two “neutral” states. On the other hand, the physical states do not carry a net color charge,but rather a superposition of states, and they must be understood as linear combinations of the elements of this matrix. With other values for $\alpha_j$, we obtain fractional values for the color charges. This theorizes the existence of exotic baryons. 

Horizontal symmetries can also be considered. For instance, in the standard electroweak model, we can impose $L^\prime_a=(U_H)_{ab}L_b$, where the flavor charges are $a=e,\mu,\tau$, which could also be summative, depending on what lepton flavor we are assigning.
\begin{equation}
Q(\textbf{6})=\left(
\begin{array}{ccc}
\tau\tau & \tau\mu & \tau e\\
 & \mu\mu & \mu e \\
 &  & ee
\end{array}
\right)\sim \begin{pmatrix}0 & -1 & +1\\ & -2 & 0 \\  &  & +2 \end{pmatrix}.
\label{flavorextet}
\end{equation} 

New composite fermions may be built with these colored sextets included in a Lagrangian with terms proportional to $\bar{S}S$ or $SSS$.

\section{Conclusions}
\label{sec:con}

In this work, we have presented a tensor-index formulation of charge operators acting on arbitrary representations of unitary gauge groups. The construction is rooted in the standard action of Cartan generators on tensor products and provides a compact bookkeeping framework for charge assignments in model building. While the underlying representation-theoretic principles are well known, the explicit operator form derived here offers a convenient and systematic implementation.

This approach opens the door to exploring new particles and theoretical models that may not be accessible through traditional methods, particularly for assigning charges to multiplets where interactions with known particles are limited or nonexistent, or to particles with unconventional charges.

Although charge quantization is not a direct consequence of the method, it can be achieved through an appropriate choice of parameters $\alpha_j$ and hypercharges \cite{PhysRevD.58.035008}. This highlights the flexibility of the approach to adapt to different physical scenarios, such as the study of gauge-coupling unification and neutrino mass generation.

In 331 models, if $Y$ is an arbitrarily small real number, say $Y\sim 10^{-3}$, the components of the sextet, octet, or decuplet with an electric charge proportional to $Y$ are of the millicharge type, while the others have exotic (but large) charges.

Finally, we highlighted that the method is not necessarily related to electric charges, as shown in subsection III D; however, it may be useful for considering other conserved quantities. For instance, in theories with color symmetries like $SU(3)_C$, the charge operator tensors may be built, demonstrating that the method still works well for these kinds of representations, taking into account an isomorphism between the color charges and electric charges.

\section*{Acknowledgment}
E.C-R. and O.P.R. are thankful for the full support and V.P. for the partial support of the Universidad Nacional de Ingeniería funding grant FC-PFR-42-2024.

\appendix

\section{Cartan generators and weight additivity}

As we stated in section \eqref{sec:ggnf}, matrix $M$ has eigenvalues equal to the weights of the fundamental multiplet.

To see the action on tensor products, let $V$ be the fundamental representation space and consider a general tensor
\begin{equation}
T \in V^{\otimes p}\otimes (V^*)^{\otimes q},
\end{equation}
written in index notation as
\begin{equation}
T^{i_1\cdots i_p}{}_{j_1\cdots j_q}.
\end{equation}

The action of a generator $X$ of the Lie algebra on a tensor product is given by the standard coproduct rule.
\begin{equation}
X \;\mapsto\; X\otimes\mathbf{1}+\mathbf{1}\otimes X,
\end{equation}
extended to all tensor factors. Moreover, in the dual representation, one has
\begin{equation}
X \mapsto -X^{T},
\end{equation}
which implies an opposite sign for lower (covariant) indices.

Consequently, the action of $X$ on the tensor $T^{i_1\cdots i_p}{}_{j_1\cdots j_q}$ is
\begin{equation}
[X(T)]^{i_1\cdots i_p}{}_{j_1\cdots j_q}=\sum_{r=1}^{p} X^{i_r}{}_{k}\,
T^{i_1\cdots k\cdots i_p}{}_{j_1\cdots j_q}-\sum_{s=1}^{q} X^{k}{}_{j_s}\,T^{i_1\cdots i_p}{}_{j_1\cdots k\cdots j_q}.
\end{equation}

Choosing $X=Q$ and using its fundamental representation matrix $M$, this expression becomes
\begin{equation}
[Q(T)]^{i_1\cdots i_p}{}_{j_1\cdots j_q}=\sum_{r=1}^{p} M^{i_r}{}_{k}\,
T^{i_1\cdots k\cdots i_p}{}_{j_1\cdots j_q}-\sum_{s=1}^{q} M^{k}{}_{j_s}\,
T^{i_1\cdots i_p}{}_{j_1\cdots k\cdots j_q}
+\frac{Y}{2}\,T^{i_1\cdots i_p}{}_{j_1\cdots j_q},
\end{equation}
which reproduces exactly the general formula given in Eq. \eqref{A_formula}.

Consequently, each tensor component corresponds to a weight state $q(i)$ and denotes the eigenvalue of the matrix $M$ acting on the fundamental basis vector.

\bibliographystyle{apsrev4-2}
\bibliography{cargaref.bib}
\end{document}